\documentclass[final]{elsart}
\usepackage{graphicx}
\usepackage{amsmath}
\usepackage{amssymb}
\usepackage{mycommands}
\usepackage{subfigure}
\usepackage{epsfig}

\journal{Physics Letters B}

\begin{document}

\begin{frontmatter}

\title{
A Precise Measurement of the Muon Neutrino-Nucleon 
Inclusive Charged Current Cross-Section off an Isoscalar 
Target in the Energy Range 
\boldmath{$2.5 < E_\nu < 40$}~GeV by NOMAD }

\author[19]{Q.~Wu}
\author[19]{S.R.~Mishra}
\author[19]{A.~Godley}
\author[19]{R.~Petti}
\author[25]{S.~Alekhin}
\author[14]{P.~Astier}
\author[8]{D.~Autiero}
\author[18]{A.~Baldisseri}
\author[13]{M.~Baldo-Ceolin}
\author[14]{M.~Banner}
\author[1]{G.~Bassompierre}
\author[9]{K.~Benslama}
\author[18]{N.~Besson}
\author[8,9]{I.~Bird}
\author[2]{B.~Blumenfeld}
\author[13]{F.~Bobisut}
\author[18]{J.~Bouchez}
\author[20]{S.~Boyd}
\author[3,24]{A.~Bueno}
\author[6]{S.~Bunyatov}
\author[8]{L.~Camilleri}
\author[10]{A.~Cardini}
\author[15]{P.W.~Cattaneo}
\author[16]{V.~Cavasinni}
\author[8,22]{A.~Cervera-Villanueva}
\author[11]{R.~Challis}
\author[6]{A.~Chukanov}
\author[13]{G.~Collazuol}
\author[8,21]{G.~Conforto \thanksref{Deceased}}
\thanks[Deceased]{Deceased}
\author[15]{C.~Conta}
\author[13]{M.~Contalbrigo}
\author[10]{R.~Cousins}
\author[9]{H.~Degaudenzi}
\author[16]{T.~Del~Prete}
\author[8,16]{A.~De~Santo}
\author[8]{L.~Di~Lella \thanksref{Now1}}
\thanks[Now1]{Now at Scuola Normale Superiore, Pisa, Italy}
\author[8]{E.~do~Couto~e~Silva}
\author[14]{J.~Dumarchez}
\author[20]{M.~Ellis}
\author[3]{G.J.~Feldman}
\author[15]{R.~Ferrari}
\author[8]{D.~Ferr\`ere}
\author[16]{V.~Flaminio}
\author[15]{M.~Fraternali}
\author[1]{J.-M.~Gaillard}
\author[8,14]{E.~Gangler}
\author[5,8]{A.~Geiser}
\author[5]{D.~Geppert}
\author[13]{D.~Gibin}
\author[8,12]{S.~Gninenko}
\author[8,22]{J.-J.~Gomez-Cadenas}
\author[18]{J.~Gosset}
\author[5]{C.~G\"o\ss ling}
\author[1]{M.~Gouan\`ere}
\author[8]{A.~Grant}
\author[7]{G.~Graziani}
\author[13]{A.~Guglielmi}
\author[18]{C.~Hagner}
\author[22]{J.~Hernando}
\author[3]{P.~Hurst}
\author[11]{N.~Hyett}
\author[7]{E.~Iacopini}
\author[9]{C.~Joseph}
\author[9]{F.~Juget}
\author[11]{N.~Kent}
\author[19]{J.J.~Kim}
\author[12]{M.~Kirsanov}
\author[6]{O.~Klimov}
\author[8]{J.~Kokkonen}
\author[12,15]{A.~Kovzelev}
\author[1,6]{A. Krasnoperov}
\author[12]{S.~Kulagin}
\author[13]{S.~Lacaprara}
\author[14]{C.~Lachaud}
\author[23]{B.~Laki\'{c}}
\author[15]{A.~Lanza}
\author[4]{L.~La Rotonda}
\author[13]{M.~Laveder}
\author[14]{A.~Letessier-Selvon}
\author[14]{J.-M.~Levy}
\author[19]{J.~Ling} 
\author[8]{L.~Linssen}
\author[23]{A.~Ljubi\v{c}i\'{c}}
\author[2]{J.~Long}
\author[7]{A.~Lupi}
\author[6]{V.~Lyubushkin}
\author[7]{A.~Marchionni}
\author[21]{F.~Martelli}\
\author[18]{X.~M\'echain}
\author[1]{J.-P.~Mendiburu}
\author[18]{J.-P.~Meyer}
\author[13]{M.~Mezzetto}
\author[11]{G.F.~Moorhead}
\author[6,7]{D.~Naumov}
\author[1]{P.~N\'ed\'elec}
\author[6]{Yu.~Nefedov}
\author[9]{C.~Nguyen-Mau}
\author[17]{D.~Orestano}
\author[17]{F.~Pastore}
\author[20]{L.S.~Peak}
\author[21]{E.~Pennacchio}
\author[1]{H.~Pessard}
\author[8]{A.~Placci}
\author[15]{G.~Polesello}
\author[5]{D.~Pollmann}
\author[12]{A.~Polyarush}
\author[11]{C.~Poulsen}
\author[6,14]{B.~Popov}
\author[13]{L.~Rebuffi}
\author[24]{J.~Rico}
\author[5]{P.~Riemann}
\author[8,16]{C.~Roda}
\author[8,24]{A.~Rubbia}
\author[15]{F.~Salvatore}
\author[6]{O.~Samoylov}
\author[14]{K.~Schahmaneche}
\author[5,8]{B.~Schmidt}
\author[5]{T.~Schmidt}
\author[13]{A.~Sconza}
\author[19]{M.~Seaton}
\author[11]{M.~Sevior}
\author[1]{D.~Sillou}
\author[8,20]{F.J.P.~Soler}
\author[9]{G.~Sozzi}
\author[2,9]{D.~Steele}
\author[8]{U.~Stiegler}
\author[23]{M.~Stip\v{c}evi\'{c}}
\author[18]{Th.~Stolarczyk}
\author[9]{M.~Tareb-Reyes}
\author[11]{G.N.~Taylor}
\author[6]{V.~Tereshchenko}
\author[12]{A.~Toropin}
\author[14]{A.-M.~Touchard}
\author[8,11]{S.N.~Tovey}
\author[9]{M.-T.~Tran}
\author[8]{E.~Tsesmelis}
\author[20]{J.~Ulrichs}
\author[9]{L.~Vacavant}
\author[4]{M.~Valdata-Nappi\thanksref{Now2}}
\thanks[Now2]{Now at Univ. of Perugia and INFN, Perugia, Italy}
\author[6,10]{V.~Valuev}
\author[14]{F.~Vannucci}
\author[20]{K.E.~Varvell}
\author[21]{M.~Veltri}
\author[15]{V.~Vercesi}
\author[8]{G.~Vidal-Sitjes}
\author[9]{J.-M.~Vieira}
\author[10]{T.~Vinogradova}
\author[3,8]{F.V.~Weber}
\author[5]{T.~Weisse}
\author[8]{F.F.~Wilson}
\author[11]{L.J.~Winton}
\author[20]{B.D.~Yabsley}
\author[18]{H.~Zaccone}
\author[5]{K.~Zuber}
\author[13]{P.~Zuccon}

\address[1]{LAPP, Annecy, France}
\address[2]{Johns Hopkins Univ., Baltimore, MD, USA}
\address[3]{Harvard Univ., Cambridge, MA, USA}
\address[4]{Univ. of Calabria and INFN, Cosenza, Italy}
\address[5]{Dortmund Univ., Dortmund, Germany}
\address[6]{JINR, Dubna, Russia}
\address[7]{Univ. of Florence and INFN,  Florence, Italy}
\address[8]{CERN, Geneva, Switzerland}
\address[9]{University of Lausanne, Lausanne, Switzerland}
\address[10]{UCLA, Los Angeles, CA, USA}
\address[11]{University of Melbourne, Melbourne, Australia}
\address[12]{Inst. for Nuclear Research, INR Moscow, Russia}
\address[13]{Univ. of Padova and INFN, Padova, Italy}
\address[14]{LPNHE, Univ. of Paris VI and VII, Paris, France}
\address[15]{Univ. of Pavia and INFN, Pavia, Italy}
\address[16]{Univ. of Pisa and INFN, Pisa, Italy}
\address[17]{Roma Tre University and INFN, Rome, Italy}
\address[18]{DAPNIA, CEA Saclay, France}
\address[19]{Univ. of South Carolina, Columbia, SC, USA}
\address[20]{Univ. of Sydney, Sydney, Australia}
\address[21]{Univ. of Urbino, Urbino, and INFN Florence, Italy}
\address[22]{IFIC, Valencia, Spain}
\address[23]{Rudjer Bo\v{s}kovi\'{c} Institute, Zagreb, Croatia}
\address[24]{ETH Z\"urich, Z\"urich, Switzerland}
\address[25]{Inst. for High Energy Physics, 142281, Protvino, Moscow, Russia}

\begin{abstract}
We present a measurement of  the muon neutrino-nucleon inclusive  
charged current cross-section, off an isoscalar target,  
in the neutrino energy range $2.5 \leq E_\nu \leq 40$~GeV.
The significance of this measurement is its  precision, 
$\pm 4$\% in $2.5 \leq E_\nu \leq 10$~GeV, and 
$\pm 2.6$\% in $10 \leq E_\nu \leq 40$~GeV 
regions,  where significant uncertainties in previous 
experiments still exist, and its  
importance to the current and proposed long baseline 
neutrino oscillation experiments.
\end{abstract}

\begin{keyword}
inclusive neutrino-nucleon cross section
\PACS 13.15.+g \sep 13.85.Lg \sep 14.60.Lm
\end{keyword}

\end{frontmatter}

\section{Motivation}
\label{section-intro}

The muon neutrino-nucleon inclusive 
charged current (\nm-N CC) cross-section
has been well measured at high neutrino 
energies  ($30 \leq E_\nu \leq 250$~GeV), 
primarily by the CCFR~\cite{ccfr} and the CDHSW~\cite{cdhsw} 
experiments. 
The average absolute 
\nm-N CC cross-section, where `N' is a nucleon in an 
isoscalar target,  above \enu\ of 30~GeV, 
$\sigma^{CC}(\nu_\mu N) = (0.677 \pm 0.014) E_\nu~cm^2/$GeV, 
is measured to a 2.1\% precision. In contrast the 
$\sigma^{CC}(\nu_\mu N)$ 
is imprecisely measured below 30~GeV.  Previous 
measurements are shown in Figure~\ref{figure-fine-cc-crs} 
and summarised in ~\cite{worldave}. 
Accurate determination of $\sigma^{CC}(\nu_\mu N)$ below 
\enu\ of 30~GeV is of interest in its own right, and offers 
insight into  CC processes such as quasi-elastic 
and resonance interactions, and their transition into 
the deep inelastic scattering region. 
The current and the proposed long baseline neutrino experiments, 
such as MINOS and NO$\nu$A at Fermilab and T2K in Japan, 
address the atmospheric $\nu$ oscillations at the 
mass-difference,
$\Delta m^2_{23} \approx 2.5 \times 10^{-3}$~eV$^2$. Given 
their typical  flight path of a few hundred kilometers, they 
use neutrino beams with energies well below 30~GeV. 
Cross sections in 
this region should be precisely known  to accurately interpret 
the results of these experiments. The NOMAD data are suitable for
such a precision $\sigma^{CC}(\nu_\mu N)$ 
measurement due to the large $\nu$-interaction 
sample, good low-energy resolution 
and a $\nu_\mu$ flux which spans 
${\cal{O}}(1) \leq E_\nu \leq 300$~GeV with a mean energy of 24.3~GeV.

\section{The Beam and the Detector}
\label{section-nomad}

The Neutrino Oscillation MAgnetic 
Detector (NOMAD) experiment at CERN used a 
neutrino beam  produced by 
the 450~GeV SPS-protons 
striking a beryllium target and producing  
secondary $\pi^{\pm}$, $K^{\pm}$, and $K^0_L$ mesons.  
The positively charged mesons were focussed by a 
system of collimators, a magnetic horn and a reflector 
into a 290~m long evacuated decay pipe. Decays of  
$\pi^{\pm}$,  $K^{\pm}$, and $K^0_L$  
produced the SPS neutrino beam. 
The average flight path of 
the neutrinos to the NOMAD was 628~m; the detector being 
836~m downstream of the Be-target.  
The SPS beamline  and the neutrino flux incident 
at NOMAD are described in~\cite{cern-beam} and~\cite{nomad-flux}.  

NOMAD was designed to search for 
\mutotau\ oscillations at $\Delta m^2 \geq 5$~eV$^2$, 
and  in this $\Delta m^2$ range it set the current best limit on this 
search~\cite{nomad-tau}. The experiment recorded over 1.7 million 
neutrino interactions in its active drift-chamber  (DC) target. 
These data are unique in that they constitute the largest
high resolution neutrino data sample with 
accurate identification of \nm, \nmb, \nel, and \neb\  
in the energy range ${\cal O}(1) \leq E_\nu \leq 300$~GeV.  
In addition, upstream of the 
active-DC target, 
the experiment recorded over 2 million  $\nu$-interactions 
in the Al-coil, and over 20 million in the Fe-scintillator 
calorimeter (FCAL).
                
The NOMAD apparatus, described in~\cite{nomad-nim}, 
was composed of several sub-detectors. The active 
target comprised 132 planes of $3 \times 3$~m$^2$ drift chambers   
with an average density similar to that of  liquid 
hydrogen (0.1~gm/cm$^3$)~\cite{nomad-dc}. 
On average, the equivalent material in the DC 
encountered by 
particles produced in a $\nu$-interaction  
was about $ 0.5~X_0$ and a quarter of an interaction 
length ($\lambda$).  
The fiducial mass of the NOMAD DC-target, 
composed  primarily 
of carbon (64\%), oxygen (22\%), nitrogen (6\%), 
and hydrogen (5\%), was 2.7 tons. 
The measured composition of the target
was 52.43\% protons and 47.57\% neutrons.
The correction for non-isoscalarity was about 5\%. 
Downstream of the DC, there were nine modules of transition radiation 
detectors (TRD), followed by a preshower (PRS) and a lead-glass 
electromagnetic calorimeter (ECAL). 
The ensemble of DC, TRD, and PRS/ECAL was placed within 
a dipole magnet providing a 0.4~T magnetic field. 
Outside  the magnet was a hadron calorimeter (HCAL),  
followed by two muon-stations comprising large area 
drift chambers separated by an iron filter. 
The two muon-stations, placed at 8- and 13-$\lambda$ downstream of 
the ECAL, provided a clean identification of the muons. 

The charged tracks in the DC were measured with an 
approximate  momentum (p)  resolution of  
$\sigma_p/p = 0.05/\sqrt{L} + 0.008p/\sqrt{L^5}$, 
$p$ in GeV and $L$ in meters,
with unambiguous charge separation in the energy range of interest. 
The $\pi^0$ component of the $\nu$-hadronic jet was measured by 
the ECAL with a resolution of $\sigma_E/E = 3.2\%/\sqrt{E}+1\%$. 
The detailed individual reconstruction 
of each charged and neutral track and their  
precise momentum vector measurement  
enabled a quantitative description of 
the event kinematics: the strength and 
basis of NOMAD analyses. In a $\nm$-CC interaction, 
in addition to the three traditional variables, 
energy ($E_\mu$), angle ($\theta_\mu$) of the  
emergent muon, and the hadron energy ($E_{HAD}$), 
the detector uniquely offered a measurement of the 
missing transverse momentum (\ptmiss) vector 
in a plane transverse to the neutrino direction. 

\section{The Analysis}
\label{section-analysis}

The $\sigma^{CC}(\nu_\mu N)$ 
was measured by dividing the fully corrected \nm-CC data by the 
corresponding \nm-flux as a function of $E_\nu$.  We first 
describe the measurement of the numerator. 
In a \nm-CC interaction, the neutrino energy ($E_\nu$) was measured 
by adding the energies of the muon (\emu) and particles 
composing the hadron-jet (\ehad)  
yielding the total visible energy ($E_{VIS}$) of the interaction.
The observed CC-data, binned in \enu\ commensurate with resolution and 
statistics, were corrected for the detector acceptance, 
the  efficiency of the cross section selection cuts, and 
the reconstruction smearing effects using \nm-CC Monte 
Carlo (MC) samples. 

To produce a clean sample of \nm-CC events, the following  
selection criteria  were imposed.
Since the $\sigma^{CC}(\nu_\mu N)$ analysis was entirely 
dominated by systematic errors, more stringent 
fiducial cuts were imposed than those used in  
statistical-error limited analyses such as~\cite{nomad-tau}. 

Next,  a successful match between a drift 
chamber track to track-segments in both muon chambers 
yielded the muon identification ($\mu$-ID).  
The polar angle of the muon with respect to the incident 
neutrino direction, $\theta_\mu$, was required to be less 
than 0.5 radians. The $P_{\mu} > 2.5$~GeV 
cut, dictated by the thickness of the HCAL preceding the
first muon station, defined the low energy limit of our measurement. 
Finally, for the 1-track sample  
a cut on the transverse muon-momentum, 
$p_t^2=(P_\mu \times \theta_\mu)^2 > 0.0025$~GeV$^2$, 
was used to eliminate the inverse muon decay events 
with minimal loss of efficiency. 
 
The standard NOMAD $\nu$-event generator, NEGLIB, 
and the detailed Monte Carlo simulation  was based upon 
LEPTO 6.1~\cite{lepto} and JETSET~\cite{jetset} generators 
for neutrino interactions and on a GEANT~\cite{geant} based 
program for the detector response. The parton content of 
the nucleon were taken from Ref.~\cite{grv}. The 
\nm-MC included deep-inelastic 
scattering (DIS), resonance (RES), and quasi-elastic (QE) processes. 
The relative abundance of 
DIS:RES:QE  samples, averaged over the 
\nm-flux,  was taken to be 1.0:0.031:0.024. 
The (QE+RES) to DIS, and QE to RES, cross sections were 
separately varied by $\pm15\%$   
and the resulting small difference in $\sigma^{CC}(\nu_\mu N)$ 
was taken as a systematic error. 
The acceptance computed using 
the total number of generated MC in the standard 
NOMAD fiducial volume~\cite{nomad-tau}   
and  the corresponding number of reconstructed 
MC events passing event selection cuts took 
into account the bias in the true average energy due to the
event reconstruction and selection process. It should be 
noted that the standard NOMAD 
fiducial volume used for generated MC 
(the denominator in acceptance calculation) was about 22\% 
larger than that used for the reconstructed sample.
A small impurity (0.7\%) due to 
neutral-current (NC), from  $\nu$ and $\overline {\nu}$ interactions, 
induced $\mu^-$-sample was corrected using the NC-MC estimation.  
The effects of the selection cuts on data and Monte Carlo are 
summarized in Table ~\ref{table-cutcc}.

\begin{table}\centering
\resizebox{\textwidth}{!}{%
\begin{tabular}{|l|c||c|c|c|c||c|c|c|c|}
\hline
Cut & Data    &  QE & RES & DIS & \nm-CC & NC        & \nel & \neb  &  \nmb  \\
\hline
Generated in Fid         &                   &        32198.8    &        42869.7    &      1364812.4    &      1439880.9    &       547103.1    &        21598.3    &         2159.9    &        35996.0   \\
Reconstructed            &      4022549.0    &        27985.2    &        37120.5    &      1182505.1    &      1247610.9    &       394053.7    &        18905.1    &         1881.3    &        31033.7   \\
Fiducial Volume          &      1815455.0    &        20265.1    &        31040.1    &      1122888.6    &      1174193.9    &       313487.8    &        18131.8    &         1547.6    &        27201.8   \\
Negative Muon            &      1069609.0    &        20114.0    &        30816.5    &       987008.8    &      1037939.3    &         6707.8    &          325.5    &           24.1    &          279.4   \\
Quality Cuts             &      1043691.0    &        19960.3    &        30527.3    &       985255.8    &      1035743.3    &         6698.7    &          325.5    &           24.1    &          279.3   \\
 $E_{\mu}>2.5$           &      1038783.0    &        19941.9    &        30509.7    &       980265.8    &      1030717.4    &         6484.5    &          316.0    &           23.2    &          270.0   \\
 $\theta_{\mu}<0.5~rad$  &      1035260.0    &        19939.4    &        30503.0    &       978387.4    &      1028829.8    &         6476.8    &          314.8    &           23.1    &          267.9   \\
    $p_t^2 > 0.0025$     &      1035107.0    &        19906.7    &        30472.9    &       978383.2    &      1028762.8    &         6476.8    &          314.8    &           23.1    &          267.9   \\
\hline
\end{tabular}}
\caption{Selection Criteria
for \nm\ Charged Current Events: The numbers of Data, 
and normalized MC samples from 
\nm-CC, NC, and \nel-, \neb-, and \nmb-CC events 
passing the  $\sigma^{CC}(\nu_\mu N)$ analysis 
cuts are shown. } 
\label{table-cutcc}
\end{table}

\section{The \nm-Flux and the Absolute Normalization}

Cross-section measurements require a knowledge of the $\nu$-flux.  
Neutrinos in the SPS beam were mainly from $\pi$, $K$, and $\mu$ decays. 
The uncertainty in modeling these secondary particles, and 
hence the $\nu$-flux, was --- 
and for all the $\sigma^{CC}(\nu_\mu N)$ measurements has been --- 
the dominant source of systematic error. 
Fortunately for NOMAD, a dedicated measurement of 
$\pi$/K yields in 450~GeV p-Be collision at various secondary 
energies and angles  was undertaken by 
the SPY experiment~\cite{spy}. The SPY measurement of 
the $\pi^{\pm}/K^{\pm}$ yields was carried out at 
discrete energies spanning 7 to 135~GeV, and a detailed 
transverse-momentum ($P_T$) scan at 15 and 40~GeV that 
were especially useful to the present measurement.  
A previous measurement of $\pi$/K  yield in a 400~GeV p-Be collision by 
Atherton $et$ $al.$~\cite{atherton} was also used in 
the $\nu$-flux determination. 
Other systematic uncertainties in the \nm-flux determination 
arose from the variation in the position of 
the primary proton beam  and the simulation of the 
propagation of secondaries through the beam line. 
The energy dependent relative \nm\ flux errors~\cite{nomad-flux} 
were the largest  source of systematic error in this analysis. 

In this analysis only the relative $\nu_\mu$-flux, i.e. 
number of \nm\ in $E_\nu$ bins, 
obtained using the SPY/Atherton measurements, was used. 
The absolute normalization of the \nm-flux was fixed   
using the world average of $\frac{\sigma (\nu N)}{E}$ 
above 40~GeV. The absolute flux normalisation was computed 
in the following energy regions: 40-100~GeV, 
40-150~GeV, 50-150~GeV, and 50-200~GeV. 
Variations in the normalisation, from these control regions, 
bracketed the error in the absolute flux normalisation process. 
In addition, the 2.1\% error in world average cross section 
was included into our error calculation.

\section{Systematic Uncertainties }

In what follows, we enumerate  sources of systematic errors 
affecting the numerator.  The muon identification-efficiency and 
energy-scale were the two most important 
measurables in the $\sigma^{CC}(\nu_\mu N)$ analysis.
First, a precise understanding of the muon-chamber efficiency and 
stability was crucial.
In a dedicated run in 1996 during the gap between the 
two neutrino spills from the SPS, 
we accumulated a large statistics of muons.  
This `Flat-top $\mu$' sample was identified 
by the veto-counter and the most upstream DCs.
The energy spectrum of the Flat-top muon sample, spanning  
4 to 50~GeV with a mean energy of 16~GeV, was similar 
to that induced by the \nm-CC events. 
The measured  absolute efficiency of the  
$\mu$-ID  for this sample was  99.96\%, in agreement
with  a detailed Monte Carlo simulation of the Flat-top muons.   
Next, we studied the stability of the $\mu$-identification by using 
the fraction of events with an identified muon,  
$\left [ \rho(\mu-ID) \right ]$, 
as a function of time spanning 1995 through 
1998,   and as a function of 15 sections of the muon chambers. 
The $\rho(\mu-ID)$ 
was stable to better than 1\% over this four-year period. 
The distribution of $\rho(\mu-ID)$, measured over  
47  running periods, was consistent with a Gaussian 
distribution with an error in the mean of 0.15\%. 
These consistency between data and MC simulation of 
$\mu$-identification ensured the accuracy of  
the \nm-CC efficiency computed by the Monte Carlo. 

In NOMAD, the $E_\mu$-scale was determined by the 
accurately measured B-field and a precise DC-alignment 
accomplished by using several 
million beam muons traversing the detector 
throughout the neutrino runs. The momentum 
scale was checked by using the invariant mass 
($M_{K_S}$) of 
over 30,000 reconstructed $K^0_S$ in the CC and NC data. 
For the $K^0_S$-momenta above 1~GeV (5~GeV), 
the data yielded 30,831 (13,765) $K^0_S$ with 
an average $M_{K_S}=498.20 \pm 0.071$~MeV 
($M_{K_S}=498.80 \pm 0.100$~MeV);  
the corresponding MC, with a +0.25\% shift in 
momentum,  yielded $498.2 \pm 0.059$~MeV 
($M_{K_S}=498.80 \pm 0.090$~MeV). The error in 
the average was estimated by RMS (=12~MeV)/$\sqrt{N}$, 
where $N$ was the number of $K^0_S$. 
In contrast, if the momentum were 
shifted by -0.5\%, the MC would yield 
$M_{K_S}=496.00 \pm 0.059$~MeV in disagreement with 
the data.  The systematic error  on the $E_\mu$-scale 
was determined to be 0.2\%.

\begin{figure}
\begin{center}
\subfigure[ 
]{
\label{fig-dtmc-ybj-eh:a}\includegraphics[width=0.44\textwidth, height=0.35\textheight]{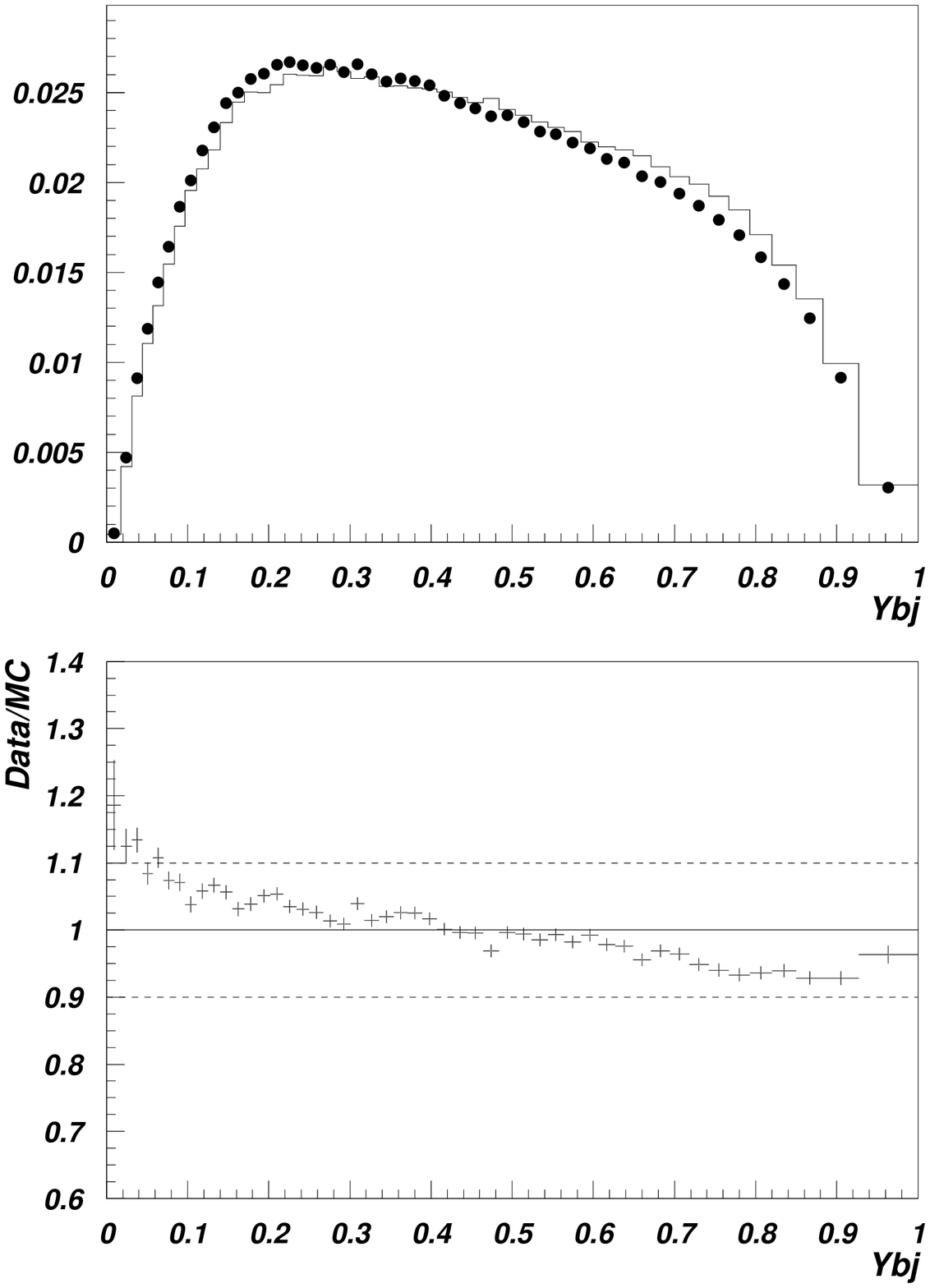}}
\hspace{0.3in}
\subfigure[ 
]{\label{fig-dtmc-ybj-eh:b}\includegraphics[width=0.44\textwidth,height=0.35\textheight]{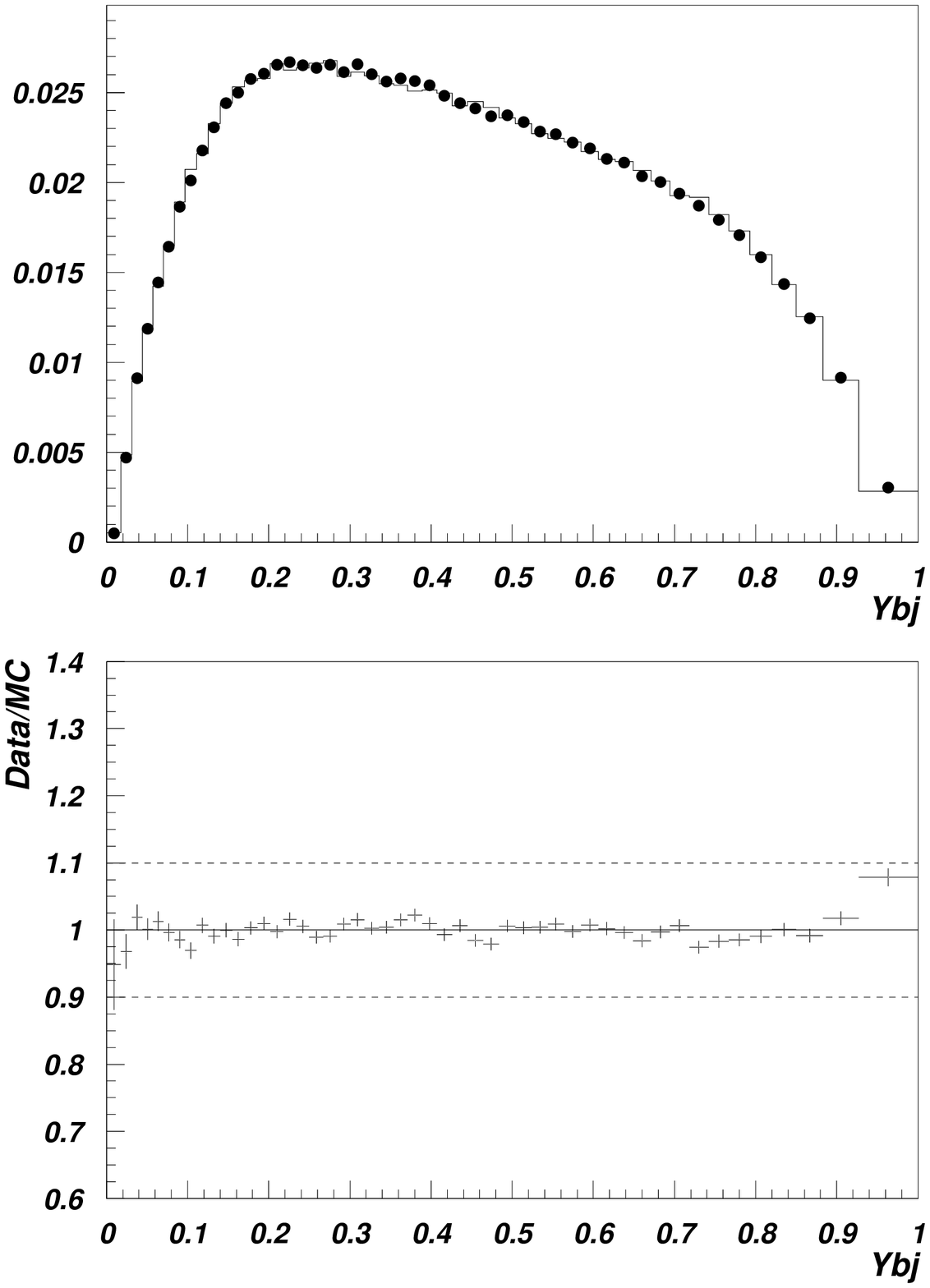}}\\

\subfigure[
]{\label{fig-dtmc-ybj-eh:c}
\includegraphics[width=0.98\textwidth,height=0.35\textheight]{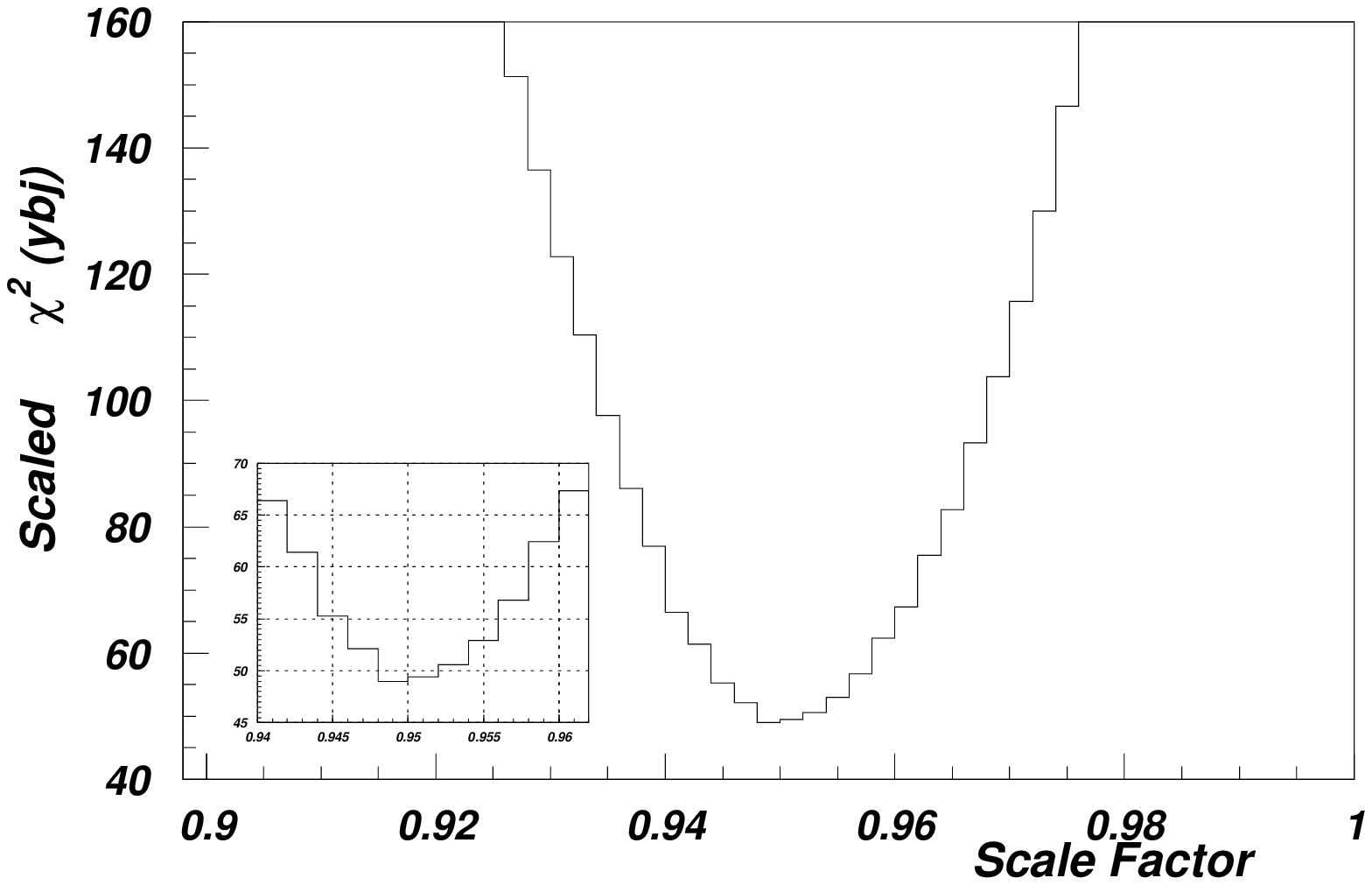}}

\caption{The Data and MC $y_{Bj}$-Distributions: 
The $y_{Bj}$ distributions for data(symbols) and MC(histogram) 
(a) before and (b) after rescaling $E_{HAD}$ are shown in the top; 
the ratio of data to Monte Carlo for the two distributions are also 
shown. The lower plot (c) shows the scaled-$\chi^2$ Distribution for $y_{BJ}$ 
as a function of \ehad-scale.}
\label{fig-dtmc-ybj-eh}
\end{center}
\end{figure}

Neutrino-induced hadron jets, including charged and neutral 
particle multiplicity and fragmentation, are poorly 
understood resulting 
in a discrepancy between the hadronic energy of 
data and MC. 
We reduced this discrepancy by correcting the simulated hadronic
energy $E_{HAD}$ by a constant factor $\kappa_H$, 
based on the distribution of 
$y_{Bj} = E_{HAD}/E_\nu = E_{HAD}/(E_{HAD}+E_\mu)$ 
in Monte Carlo and data. 
We relied on the precise measurement of $E_\mu$.
To determine the  $\kappa_H$  trials were made to minimize  
the $\chi^2$ between data and MC $y_{Bj}$- and 
$E_{HAD}$-distributions, for events with $E_{HAD} \geq 2.5$~GeV, 
by varying $\kappa_H$ from 0.9 to 1.1 in steps of 0.002 in the MC. 
The $\chi^2$  was minimised at  $\kappa_H$ of 0.950, 
$i.e.$ the MC overestimated  \ehad\ by 5\%. 
The comparison of the  $y_{Bj}$ distribution 
between data and the uncorrected-MC 
is shown in Figure~\ref{fig-dtmc-ybj-eh:a},  
where $\frac{\chi^2}{DoF}$ is $795.1/49$. 
The corresponding comparison after 
correcting the MC-$E_{HAD}$ is shown in 
Figure~\ref{fig-dtmc-ybj-eh:b},  where $\frac{\chi^2}{DoF}$ is $89.6/49$. 
To determine the error on $\kappa_H$ we formed a `scaled'-$\chi^2$ 
which yielded the scaled-$\frac{\chi^2}{DoF}$ equal to 
unity at $\kappa_H$ of 0.950. This was achieved 
by increasing the errors by 40\%.
Figure~\ref{fig-dtmc-ybj-eh:c} shows the scaled-$\chi^2$ as 
a function of $\kappa_H$. 
An increase of 1.0 from the minimum in the 
the scaled-$\chi^2$ (see the inset)  was used to set 
the uncertainty on the optimum 
$\kappa_H$ value of $0.950$. Additionally, the fiducial and 
kinematic cuts were varied and the 
range in $\kappa_H$ was redetermined  
for unity variation in the scaled-$\chi^2$. We concluded  that 
an error of $\pm 0.006$ bracketed the error on 
$\kappa_H$.  Since $\kappa_H$ 
was determined over the entire range of $E_\nu$, to cover 
possible variations in $\kappa_H$ as a function 
of $E_\nu$, we increased the scale-error by 50\%. 
Correcting $E_{HAD}$ in the MC by $\kappa_H$ also
improved the agreement between the 
data and MC distributions of other kinematic 
variables: $Q^2$, $W^2$, and $x_{Bj}$ where the improvement 
was comparable to that shown in Figure~\ref{fig-dtmc-ybj-eh:b}.
The $E_{HAD}$ correction factor determined  
in this analysis is closer 
to unity than the value of 0.93 used in our previous 
analyses~\cite{nomad-nmne} 
because  of better tuning of the Monte Carlo and 
a reprocessing of the data  that improved the reconstruction 
of high multiplicity events. The difference in the 
$\sigma^{CC}(\nu_\mu N)$ due to the 
$\pm 0.009$ uncertainty on $\kappa_H$ 
was computed and assigned as the 
systematic error. This systematic uncertainty would have
to be a factor of 2.5 times larger to make it 
one of the dominant systematic errors in the analysis. 
Although the 0.9\% error 
in the $E_{HAD}$-scale is adequate for the present 
inclusive $\sigma^{CC}(\nu_\mu N)$ measurement, efforts 
are underway to reduce this error to the 0.5\% level  using 
improved modeling~\cite{nuage} and analysis for the future 
\nm-CC differential cross-section as a function of $E_\nu$, $x_{Bj}$, 
and $y_{Bj}$, and the weak mixing angle measurements. 
Table~\ref{table-errors} lists the systematic errors on 
the $\sigma^{CC}(\nu_\mu N)/E_\nu$  as a function of visible energy.

Radiative corrections~\cite{bardin} that affected measurables, 
such as $E_\mu$, $\theta_\mu$, and $E_{HAD}$,  were folded into 
the $\sigma^{CC}(\nu_\mu N)$ measurement as a function of  
$E_\nu$. The dominant radiative effect, typically 
less than 1\% on $\sigma /E_\nu$, occurred when a photon, 
radiated by the muon, was measured as part of the hadronic 
system. No other effort was made to correct the \nm-CC cross section to the 
Born-level.  

\begin{table}
 \begin{tabular}{|c|c|c|c|c|c|}
\hline
 $E_{vis}$& Relative & Normalisation  & ${\mu}$-Acceptance & $E_{HAD}$-Scale & QE:RES:DIS  \\
(GeV)      & Flux       & Region             &                                   &                              &     \\
\hline

2.5--10  &      0.026  &      0.005  &      0.004  &      0.008  &    0.002\\
10--15   &      0.018  &      0.004  &      0.001  &      0.004  &    0.001\\
15--30   &      0.016  &      0.005  &      0.001  &      0.006  &    0.000\\
30--50   &      0.022  &      0.005  &      0.000  &      0.003  &    0.000\\
50--100  &     0.040  &      0.004  &      0.000  &    0.005    &    0.000\\
100--300 &    0.051  &      0.004  &      0.002  &   0.010     &    0.000\\
\hline
\end{tabular}
\caption{Systematic Uncertainties on $\sigma/E$ in $E_\nu$-bins.}
\label{table-errors}
\end{table}

\section{Result}

\begin{figure}
\begin{center}
\includegraphics[scale=0.7]{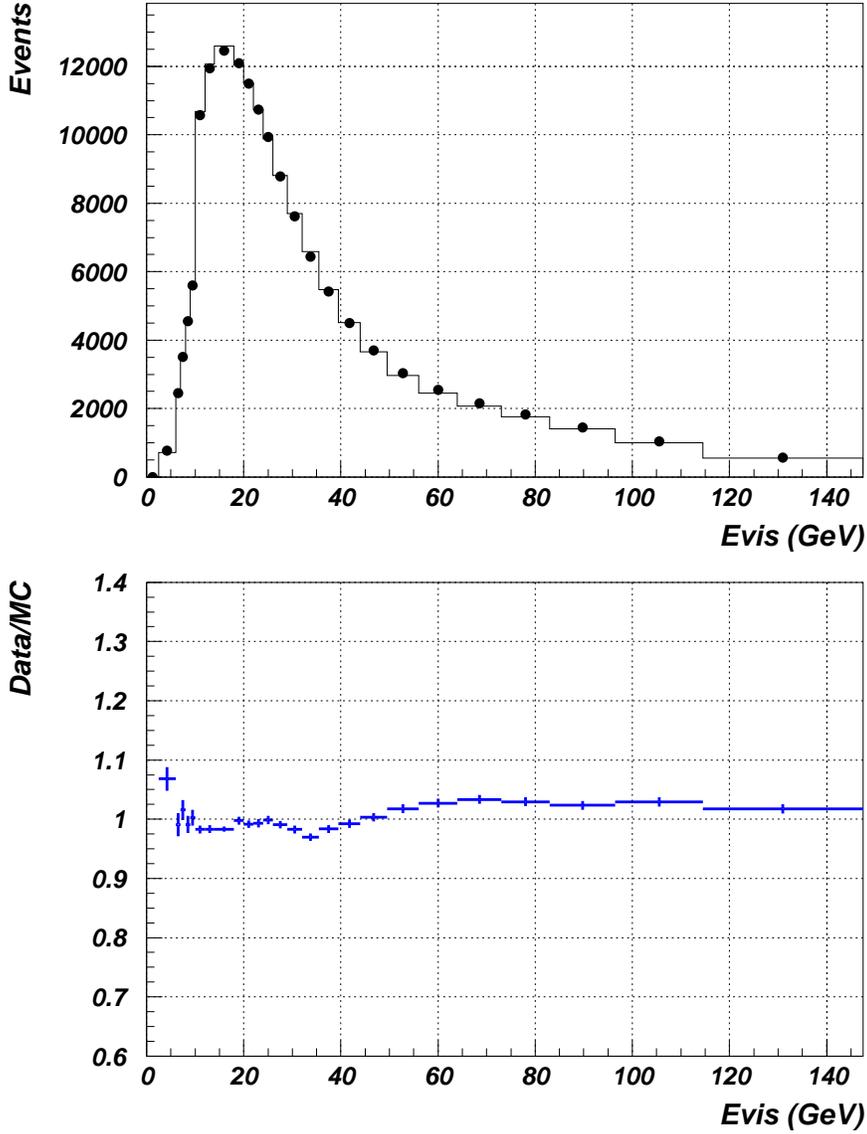}
\caption{Distributions of $E_{VIS}$ for
Data(symbols) and Monte Carlo (histogram): The 
$E_{HAD}$ correction is applied to the MC. Only the 
statistical errors are shown. 
The ratio of data to Monte Carlo is  also presented.}
\label{fig-dtmc-evis}
\end{center}
\end{figure}

After the $E_{HAD}$-scale correction,
we present the  $E_{VIS}$ comparison between
data and MC in Figure~\ref{fig-dtmc-evis}. 
Except for the lowest energy bin, the agreement
is better than  2\% in the  energy range shown. 
We point out that the \nm-CC cross-section 
was not modified in the Monte Carlo. 
The inclusive \nm-CC cross-sections 
were derived from this distribution. 
The final result of the measurement of the inclusive \nm\ charged 
current (CC) cross section is summarized in Table~\ref{table-xsec}. 
The \enu-bin, the average-\enu, 
number of observed data and background (mainly from NC)
events passing the selection criteria are listed 
respectively in the first four columns. 
The observed data are corrected by subtracting the 
background, and then dividing  by the efficiency (5-column). 
The cross section, after correcting for non-isoscalarity, 
was calculated by dividing the corrected data (6-column) 
by the flux  after absolute normalisation (7-column) 
and the average-\enu. The $\sigma^{CC}(\nu_\mu N)/E_\nu$ 
with the statistical, systematic, and total errors 
are shown in the last four columns of Table~\ref{table-xsec}. 

\begin{table}\centering
\resizebox{0.85\textwidth}{!}{%
\begin{tabular}{|c|c|c|c|c|c|c|c|c|c|c|}
\hline
$E_{\nu}-Bin$ & Avg.$E_\nu$ & Data & Bkgd & Eff. &Cor.Data  & Flux        & 
$\sigma /E_\nu$        & Stat               & Syst  & Total\\
    (GeV)         & (GeV)            &          &           &       &                & ($10^5$) & 
                       &  Err.               & Err.   &  Err.\\
\hline
  2.5-  6.0    &    4.60         &        5429.0   &          51.2   &         0.409   &       13296.9   &          4.07   &     0.786   &         0.011   &         0.035   &         0.037  \\
  6.0-  7.0    &    6.50         &        4917.0   &          45.0   &         0.452   &       10778.5   &          2.40   &     0.763   &         0.011   &         0.036   &         0.038   \\
  7.0-  8.0    &    7.50         &        7011.0   &          53.2   &         0.445   &       15625.2   &          3.20   &     0.722   &         0.009   &         0.035   &         0.036   \\
  8.0-  9.0    &    8.50         &        9119.0   &          46.0   &         0.445   &       20369.0   &          3.79   &     0.701   &         0.007   &         0.033   &         0.034   \\
  9.0- 10.0    &    9.50         &       11192.0   &          50.5   &         0.443   &       25171.9   &          4.10   &   0.716   &         0.007   &         0.033   &         0.034   \\
 10.0- 11.0    &   10.50         &       20244.0   &          87.9   &         0.704   &       28629.3   &          4.29   &  0.706   &         0.005   &         0.026   &         0.026   \\
 11.0- 12.0    &   11.50         &       22051.0   &          91.2   &         0.698   &       31471.8   &          4.31   &  0.705   &         0.005   &         0.024   &         0.025   \\
 12.0- 13.0    &   12.50         &       23349.0   &         100.7   &         0.685   &       33936.8   &          4.33   &  0.697   &         0.005   &         0.024   &         0.025   \\
 13.0- 14.0    &   13.50         &       24433.0   &          94.3   &         0.686   &       35462.1   &          4.17   &   0.700   &         0.005   &         0.024   &         0.025   \\
 14.0- 15.0    &   14.50         &       24802.0   &          91.1   &         0.682   &       36249.3   &          3.98   &   0.698   &         0.004   &         0.025   &         0.025   \\
 15.0- 17.5    &   16.20         &       62447.0   &         249.7   &         0.678   &       91750.9   &          9.00   &  0.698   &         0.003   &         0.025   &         0.025   \\
 17.5- 20.0    &   18.70         &       60825.0   &         246.5   &         0.686   &       88315.5   &          7.48   &  0.700   &         0.003   &         0.025   &         0.025   \\
 20.0- 22.5    &   21.20         &       57249.0   &         240.2   &         0.690   &       82590.0   &          6.18   &  0.699   &         0.003   &         0.024   &         0.024   \\
 22.5- 25.0    &   23.70         &       51919.0   &         226.6   &         0.691   &       74772.6   &          5.04   &  0.694   &         0.003   &         0.024   &         0.024   \\
 25.0- 27.5    &   26.20         &       46696.0   &         233.4   &         0.693   &       67054.3   &          4.09   &  0.694   &         0.003   &         0.025   &         0.025   \\
 27.5- 30.0    &   28.70         &       41462.0   &         239.3   &         0.696   &       59235.3   &          3.30   &  0.694   &         0.003   &         0.025   &         0.025   \\
 30.0- 35.0    &   32.30         &       68858.0   &         431.4   &         0.708   &       94730.8   &          4.91   &  0.677   &         0.003   &         0.026   &         0.026   \\
 35.0- 40.0    &   37.30         &       54059.0   &         420.5   &         0.704   &       75291.1   &          3.33   &  0.681   &         0.003   &         0.026   &         0.026   \\
 40.0- 45.0    &   42.40         &       43650.0   &         379.9   &         0.715   &       61212.5   &          2.35   &  0.675   &         0.003   &         0.028   &         0.028   \\
 45.0- 50.0    &   47.40         &       36135.0   &         326.3   &         0.718  &       49084.9   &          1.71   &   0.682   &         0.004   &         0.027   &         0.027   \\
 50.0- 60.0    &   54.60         &       57357.0   &         618.2   &         0.733   &       77653.8   &          2.35   &  0.670   &         0.003   &         0.028   &         0.028   \\
 60.0- 70.0    &   64.70         &       45880.0   &         509.8   &         0.733   &       61753.1   &          1.57   &   0.675   &         0.003   &         0.031   &         0.031   \\
 70.0- 80.0    &   74.80         &       38523.0   &         409.6   &         0.700   &       54226.5   &          1.18   &   0.684   &         0.003   &         0.037   &         0.037   \\
 80.0- 90.0    &   84.80         &       32054.0   &         309.1   &         0.666   &       47043.6   &          0.92   &   0.678   &         0.004   &         0.041   &         0.041   \\
 90.0-100.0    &   94.80         &       25884.0   &         231.8   &         0.636   &       39517.5   &          0.70   &  0.677   &         0.004   &         0.043   &         0.043   \\
100.0-115.0    &  107.00         &       29673.0   &         258.4   &         0.628   &       46821.5   &          0.72   &  0.674   &         0.004   &         0.048   &         0.048   \\
115.0-130.0    &  122.00         &       20327.0   &         176.7   &         0.608   &       32923.4   &          0.46   &  0.661   &         0.005   &         0.048   &         0.048   \\
130.0-145.0    &  136.90         &       14204.0   &         117.7   &         0.583   &       24337.2   &          0.29   &  0.671   &         0.006   &         0.054   &         0.054   \\
145.0-200.0    &  165.90         &       24007.0   &         170.9   &         0.545   &       43805.8   &          0.44   &  0.667   &         0.004   &         0.054   &         0.054   \\
200.0-300.0    &  228.30         &        8589.0   &          56.0   &         0.496   &       17183.5   &          0.12   &    0.721   &         0.008   &         0.060   &         0.061   \\
\hline
\end{tabular}}
\caption{Summary of the $\nm$-CC Cross Section, 
$\sigma(10^{-38}cm^2)/E(GeV)$, Analysis: The fifth-column represents 
the efficiency folded with the acceptance, 
see Section~\ref{section-analysis}. The $\sigma/E_\nu$ is 
presented for an iso-scalar nucleon within the NOMAD target.}
\label{table-xsec}
\end{table}

\begin{figure}
\begin{center}
\includegraphics[scale=0.7]{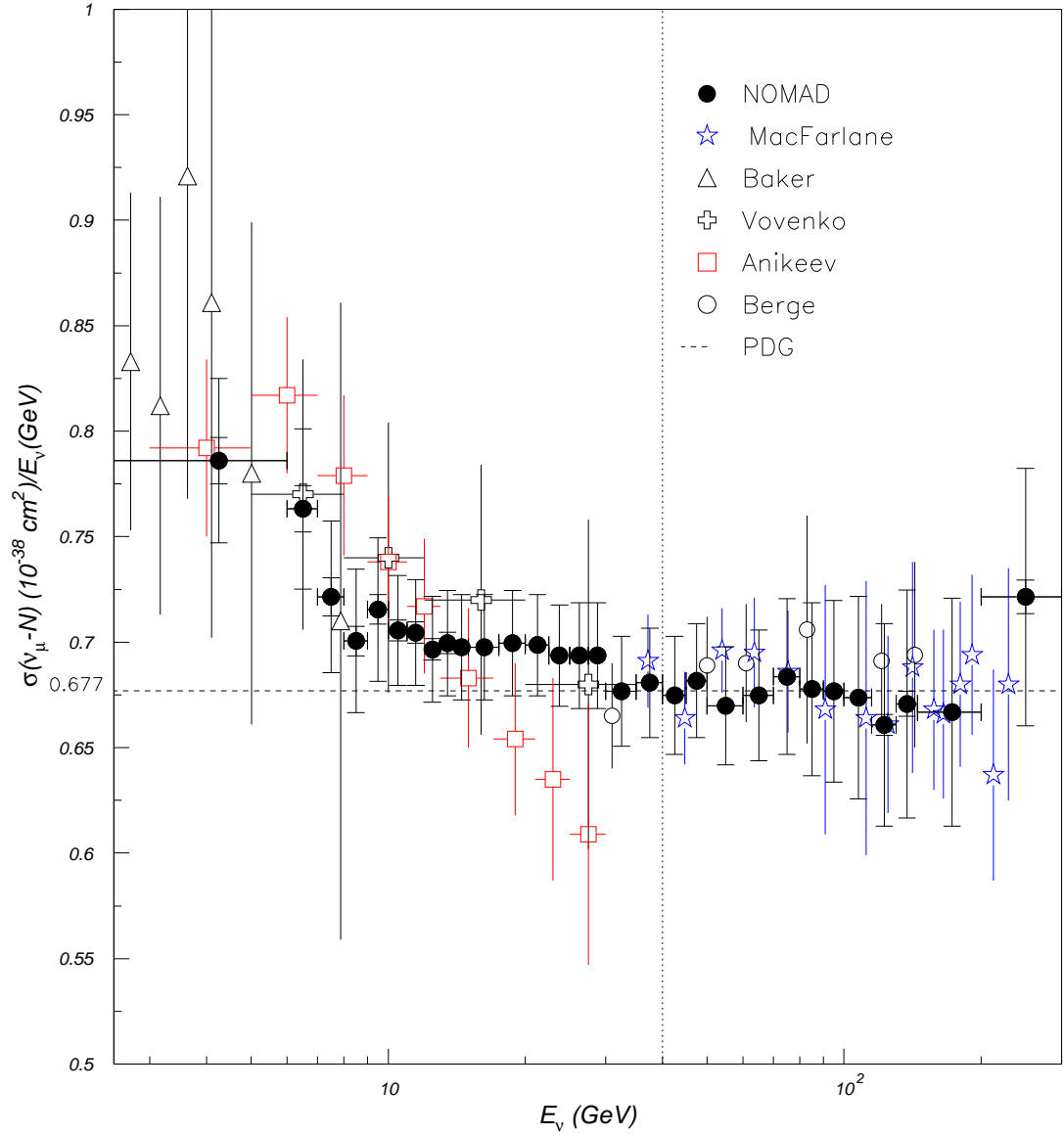}
\caption{Inclusive \nm-N Charge Current Cross Section -vs- \enu: 
The $\sigma^{CC}(\nu_\mu N)$/\enu\ is plotted as a function of 
\enu, where $N$ represents an iso-scalar nucleon 
within the the NOMAD target. The outer (inner) 
error bars show the total (statistical) error. 
Other measurements in this plot are by D.B.MacFarlane 
$et$ $al.$\cite{ccfr},
J.P.~Berge $et$ $al.$\cite{cdhsw}, 
N.J.Baker $et$ $al.$\cite{baker}, A.S. Vovenko $et$ $al.$\cite{vovenko}, and  
V.Anikeev $et$ $al.$\cite{anikeev}.  
The region $E_\nu \geq 40$~GeV was 
used to normalize the $\sigma^{CC}(\nu_\mu N)$/\enu\  
to the asymptotic world average~\cite{worldave},
shown as the dashed line, derived from high energy data.}
\label{figure-fine-cc-crs}
\end{center}
\end{figure}

The inclusive $\nm$ CC cross section divided by \enu\ is plotted  
as a function of \enu\  in Figure~\ref{figure-fine-cc-crs} 
together with existing 
measurements.  From this plot,  agreement with the existing data 
above $E_\nu \geq 30$~GeV 
is seen: $\sigma^{CC}(\nu_\mu N)/\enu$ is flat above 30~GeV; 
it rises at lower energies  due to the increasing presence of the 
non-scaling processes. In the sub 30~GeV
region, the NOMAD measurements  
improve the precision. We note that in earlier publications 
on $\sigma^{CC}(\nu_\mu N)$,  
in the $2 \leq E_\nu \leq 30$~GeV region, 
such as by Baker $et$ $al.$~\cite{baker} and 
Anikeev $et$ $al.$~\cite{anikeev},  
the $\nu_\mu$-flux  was constrained using QE events by 
selecting low-$E_{HAD}$ events. The proponents then 
used the QE cross-section to deduce the flux, assuming 
that the QE cross-section was known 
to a $\pm 5\%$ precision. This, in our opinion, was an 
optimistic precision. A compilation of all the QE-measurements 
shows that the error on the QE cross-section, in the 
$2 \leq E_\nu \leq 30$~GeV range, is close 
to 15\% as currently used by NOMAD in this paper and 
MINOS~\cite{minos} collaborations.

\section*{Acknowledgments}
We gratefully acknowledge the CERN SPS staff for the magnificient 
performance of the neutrino beam. The experiment was supported 
by the following agencies: 
ARC and DEST of Australia; IN2P3 and CEA of France, BMBF of 
Germany, INFN of Italy, JINR and INR of Russia, FNSRS of 
Switzerland, DOE, NSF, Sloan, and Cottrell Foundations of 
USA.

\end{document}